\documentclass[english,aip, preprint]{revtex4-1}
\usepackage[T1]{fontenc}
\usepackage[latin9]{inputenc}
\usepackage{color}
\usepackage{bm}
\usepackage{amsmath}
\usepackage{graphicx}
\usepackage{esint}
\usepackage{subscript}

\makeatletter

\providecommand{\tabularnewline}{\\}

 \@ifundefined{textcolor}{}
 {%
   \definecolor{BLACK}{gray}{0}
   \definecolor{WHITE}{gray}{1}
   \definecolor{RED}{rgb}{1,0,0}
   \definecolor{GREEN}{rgb}{0,1,0}
   \definecolor{BLUE}{rgb}{0,0,1}
   \definecolor{CYAN}{cmyk}{1,0,0,0}
   \definecolor{MAGENTA}{cmyk}{0,1,0,0}
   \definecolor{YELLOW}{cmyk}{0,0,1,0}
 }

\usepackage{bm}

\@ifundefined{showcaptionsetup}{}{%
 \PassOptionsToPackage{caption=false}{subfig}}
\usepackage{subfig}
\makeatother

\usepackage{babel}
\begin{document}

\title{Equivalence of Particle-Particle Random Phase Approximation Correlation
Energy and Ladder-Coupled-Cluster-Double}

\author{Degao Peng}

\affiliation{Department of Chemistry, Duke University, Durham, North Carolina,
United States, 27708}

\author{Stephan N. Steinmann}

\affiliation{Department of Chemistry, Duke University, Durham, North Carolina,
United States, 27708}

\author{Helen van Aggelen}

\affiliation{Department of Chemistry, Duke University, Durham, North Carolina,
United States, 27708}

\affiliation{Ghent University, Department of Inorganic and Physical Chemistry,
9000 Ghent, Belgium}

\author{Weitao Yang}

\email{weitao.yang@duke.edu}

\affiliation{Department of Chemistry, Duke University, Durham, North Carolina,
United States, 27708}

\affiliation{Department of Physics, Duke University, Durham, North Carolina, United
States, 27708}
\begin{abstract}
We present an analytical proof and numerical demonstrations of the
equivalence of the correlation energy from particle-particle random
phase approximation (pp-RPA) and ladder-couple-cluster-doubles (ladder-CCD).
These two theories reduce to the identical algebraic matrix equation
and correlation energy expressions, under the assumption that the
pp-RPA equation is stable. The numerical examples illustrate that
the correlation energy missed by pp-RPA in comparison with couple-cluster
single and double is largely canceled out when considering reaction
energies. This theoretical connection will be beneficial to future
pp-RPA studies based on the well established couple cluster theory.
\end{abstract}

\keywords{Particle-particle random phase approximation, ladder-couple-cluster-double,
ladder diagrams, correlation energy}

\maketitle

\section{Introduction}

The random-phase approximation (RPA) was originally proposed back
in the 1950s by Pine and Bohm\cite{Bohm51625,Pines52338} to treat
the homogeneous electron gas. Since then, the idea of RPA has spawned
the studies of excitation energies, linear-response functions and
correlation energies in solid state physics\cite{Lindhard548,Eguiluz831907,Langreth751425,Giuliani05},
nuclear physics\cite{Blaizot86,Ring04,Thouless60225,Thouless6178,Thouless62211,Marshalek69569},
and quantum chemistry\cite{Eshuis10234114,Furche08114105,Kurth9910461,Furche01195120}.
In the recent decade, there is a renaissance of interest in the RPA
correlation energy in molecular science because of its correct description
of van der Waals interaction\cite{Furche01195120}, the correct dissociation
limit of H$_{2}$\cite{Fuchs05094116} and its perspective of the
adiabatic connection in density-functional theory (DFT)\cite{Furche01195120},
with relatively low scaling ($O(N^{4}\log N)$ by Eshuis \emph{et
al.}\cite{Eshuis12084105} and $O(N^{4})$ by Ren \emph{et al.}\cite{Ren127447}
with $N$ the number of basis functions). Correlation energy studies
beyond RPA is an active field of research that achieves exciting results\cite{Zhu10244108,Toulouse2010,Hesselmann112473,Mori-Sanchez12042507,Hellgren12034106,Furche05164106,Gould13014109}.

Usually, RPA describes exclusively the particle-hole channel of correlations
in molecular science. In nuclear physics, however, the particle-particle
channel of RPA (pp-RPA) is also widely discussed\cite{Ring04,Blaizot86,Toivanen95410,Fukuda64A932,Mulhall67261,Rowe681283,Rowe68153,Ripka69489,Vary71479,Blanchon10034313,Pacheco02044004}.
In chemistry, the pp-RPA has only been used in computational study
of Auger spectroscopy which involves double ionization of molecules\cite{Liegener962940,Liegener82188}.
The application of pp-RPA to calculate the correlation energy for
molecular systems is absent until the recent work of van Aggelen \emph{et
al.} on pp-RPA\cite{vanAggelen13}, which shows promising results
in describing systems with both fractional charge and fractional spin.
Furthermore, Ref. \cite{vanAggelen13} establishes an adiabatic connection
for the exchange-correlation energy in terms of the dynamic paring
matrix \textcolor{black}{fluctuation}, parallel to the adiabatic connection
fluctuation dissipation (ACFD) theorem in terms of the density fluctuation\cite{Gunnarsson764274,Langreth751425}.
Like the ACFD theorem, this new adiabatic connection is in principle
exact, but requires the particle-particle propagator as a function
of the interaction strength. The pp-RPA has been shown to be the first-order
approximation to the paring matrix fluctuation. To distinguish the
two RPAs of different channels, we will, hereafter, refer to the conventional
particle-hole RPA as ph-RPA. 

According to Scuseria \emph{et al}.\cite{Scuseria08231101}, the ph-RPA
correlation energy is equivalent to a direct ring coupled cluster
double (direct-ring-CCD). We now prove that pp-RPA is equivalent to
ladder-CCD, assuming that the pp-RPA equation of the system is stable.
The pp-RPA correlation energy can be interpreted as the sum of all
ladder diagrams\cite{Blaizot86} or zero-point pairing vibrational
energy beyond the mean-field approximation\cite{Ring04}. The pp-RPA
wavefunction of an exponential form has been proposed\cite{Ring04}
with the argument of Thouless theorem\cite{Thouless60225} under the
quasi-boson approximation. However, its ladder-CCD nature has never
been explicitly stated in the literature. The establishment of the
equivalence of pp-RPA and ladder-CCD will be beneficial to study pp-RPA
properties. Furthermore, in the coupled cluster framework, the excited
states based on the pp-RPA wavefunction can be strategically obtained
via equation-of-motion coupled-cluster\cite{Nooijen1995,Nooijen1997,Gwaltney1999,Shavitt09}
or, equivalently, linear-response coupled-cluster theory\cite{Sekino84255,Kowalski07164105}.

\section{The pp-RPA equation and its stability\label{sec:The-ppRPA-equation}}

The pp-RPA equation can be derived from the two-particle Green's function,
the equation-of-motion ansatz, or the linear-response time-dependent
Hartree-Fock-Bogoliubov approximation (TDHFB)\cite{Blaizot86,Ring04,vanAggelen13}.
The resulting generalized eigenvalue equation is very similar to the
ph-RPA equation (see, for example, Ref. \cite{Blaizot86,Ring04,Furche01195120,Scuseria08231101}
for the ph-RPA equation),
\begin{equation}
\left[\begin{array}{cc}
\mathbf{A} & \mathbf{B}\\
\mathbf{B}^{\dagger} & \mathbf{C}
\end{array}\right]\left[\begin{array}{c}
\mathbf{x}_{n}\\
\mathbf{y}_{n}
\end{array}\right]=\omega_{n}\left[\begin{array}{cc}
\mathbf{I} & \mathbf{0}\\
\mathbf{0} & -\mathbf{I}
\end{array}\right]\left[\begin{array}{c}
\mathbf{x}_{n}\\
\mathbf{y}_{n}
\end{array}\right],\label{eq:ppRPA}
\end{equation}
where
\begin{equation}
A_{ab,cd}=(\epsilon_{c}+\epsilon_{d}-2\nu)\delta_{ac}\delta_{bd}+\langle ab||cd\rangle,\label{eq:Adef}
\end{equation}
\begin{equation}
C_{ij,kl}=-(\epsilon_{k}+\epsilon_{l}-2\nu)\delta_{ki}\delta_{jl}+\langle ij||kl\rangle,\label{eq:Cdef}
\end{equation}
and
\begin{equation}
B_{ab,ij}=\langle ab||ij\rangle.\label{eq:Bdef}
\end{equation}
We use indexes $i$, $j$, $k$, $l\ldots$ for occupied spin orbitals
(holes), $a$, $b$, $c$, $d\ldots$ for unoccupied spin orbitals
(particles), and $u$, $v$, $s$, $t\ldots$ for general spin orbitals.
Furthermore, $m$, $n$ are used to denote  eigenvector and eigenvalue
indexes. Additionally, $\epsilon_{u}$ is the molecular orbital eigenvalue,
and $\langle uv||st\rangle$ is the antisymmetrized two-electron integral
\begin{equation}
\langle uv||st\rangle=\langle uv|st\rangle-\langle uv|ts\rangle,\label{eq:ERI}
\end{equation}
where
\begin{equation}
\langle uv|st\rangle=\sum_{\sigma_{1}\sigma_{2}}\int d\mathbf{r}_{1}d\mathbf{r}_{2}\frac{\phi_{u}^{*}(\mathbf{r}_{1}\sigma_{1})\phi_{v}^{*}(\mathbf{r}_{2}\sigma_{2})\phi_{s}(\mathbf{r}_{1}\sigma_{1})\phi_{t}(\mathbf{r}_{2}\sigma_{2})}{|\mathbf{r}_{1}-\mathbf{r}_{2}|}.\label{eq:ERI1}
\end{equation}
The chemical potential $\nu$ is not an necessity in the equation-of-motion
derivation\cite{Ring04}; while during the derivation from the two-particle
Green's function and the TDHFB\cite{Blaizot86,vanAggelen13}, $\nu$
is used to ensure that the ground state has the desired number of
electrons $N$. In practice, it is usually approximated to be half
of HOMO (highest occupied molecular orbital) and LUMO (lowest unoccupied
molecular orbital) eigenvalues\cite{vanAggelen13}. We will later
show that the exact choice of the chemical potential is unimportant
within a certain range as long as the pp-RPA equation is stable.

The indexes of the matrix are either hole pairs or particle pairs,
but no particle-hole pairs. These indexes have only $i>j$ for hole
pairs and $a>b$ for particle pairs to eliminate the redundancy. The
number of particle (hole) pairs is
\begin{equation}
N_{pp(hh)}=\frac{1}{2}N_{\text{vir(occ)}}(N_{\text{vir(occ)}}-1),\label{eq:dimension-1}
\end{equation}
where $N_{\text{vir(occ)}}$ is the number of virtual (occupied) orbitals.
In general, $N_{pp}$ is much larger than $N_{hh}$. The dimension
of the upper left (lower right) identity matrix in Eq. (\ref{eq:ppRPA})
is $N_{pp}\times N_{pp}$ ($N_{hh}\times N_{hh}$), the same dimension
of $\mathbf{A}$ ($\mathbf{C}$). For the rest of the paper, the dimensions
of identity matrices will be omitted as they are clear from the context.
The difference of the dimensions of $\mathbf{A}$ and $\mathbf{C}$
makes the pp-RPA equation quite different from the usual ph-RPA equation
or the linear-response time-dependent density-functional theory equation\cite{Casida95155}. 

For simplicity, we use a compact matrix notation
\begin{equation}
\mathbf{Mz}_{n}=\omega_{n}\mathbf{Wz}_{n},\label{eq:compact}
\end{equation}
to denote Eq. (\ref{eq:ppRPA}), where $\mathbf{M}$ is the Hermitian
matrix on the left hand side 
\begin{equation}
\mathbf{M}=\left[\begin{array}{cc}
\mathbf{A} & \mathbf{B}\\
\mathbf{B}^{\dagger} & \mathbf{C}
\end{array}\right],\label{eq:Mdef}
\end{equation}
$\mathbf{W}$ is the non-positive definite metric
\begin{equation}
\mathbf{W}=\left[\begin{array}{cc}
\mathbf{I} & \mathbf{0}\\
\mathbf{0} & -\mathbf{I}
\end{array}\right],\label{eq:Wdef}
\end{equation}
and $\mathbf{z}_{n}$ is the full eigenvector
\begin{equation}
\mathbf{z}_{n}=\left[\begin{array}{c}
\mathbf{x}_{n}\\
\mathbf{y}_{n}
\end{array}\right],\label{eq:zdef}
\end{equation}
with its eigenvalue $\omega_{n}$. Due to the non-positive definite
metric $\mathbf{W}$, Eq. (\ref{eq:ppRPA}) is not guaranteed to have
all real eigenvalues. We call $\mathbf{z}_{n}^{\dagger}\mathbf{Wz}_{n}$
the signature of an eigenvector $\mathbf{z}_{n}$. The signature can
be positive, zero, or negative. The zero signature coincides with
an imaginary eigenvalue (see Subsection \ref{sec:The-zero-signature}
in Appendix), while positive and negative signatures are associated
with real eigenvalues. We categorize the eigenvectors according to
their signature, where eigenvectors with positive signatures are called
$N+2$ excitations and eigenvectors with negative signatures are called
$N-2$ excitations. For a diagonalizable pp-RPA equation with all
real eigenvalues, according to Subsection \ref{sec:orthonormalization}
in Appendix, the orthonormalization of the eigenvectors can be written
as, 
\begin{equation}
\mathbf{Z}^{\dagger}\mathbf{WZ}=\mathbf{W},\label{eq:eigenvectorNormalization}
\end{equation}
with all $N+2$ eigenvectors to the left of all $N-2$ eigenvectors
in $\mathbf{Z}$. This special arrangement will be kept all through
the paper.

When all the eigenvalues of a diagonalizable pp-RPA equation are real,
the pp-RPA equation is defined to be stable if all the $N+2$ excitation
eigenvalues are positive and $N-2$ excitation eigenvalues are negative,
i.e. $\min_{n}\omega_{n}^{N+2}>0>\max_{m}\omega_{m}^{N-2}$. With
the eigenvector arrangement according to signatures, the stability
condition can be expressed in a concise equation,
\begin{equation}
\text{sign}(\bm{\omega})=\mathbf{W},\label{eq:stability-1}
\end{equation}
where $\text{sign}(\bm{\omega})$ is the sign function\cite{Higham08}
of the eigenvalue matrix $\bm{\omega}$, which gives $[\text{\text{sign}}(\bm{\omega})]_{nm}=\delta_{nm}\text{sign}(\omega_{n})$
since $\bm{\omega}$ is diagonal. Note that Eq. (\ref{eq:eigenvectorNormalization})
is a necessary but not sufficient condition for the stability of Eq.
(\ref{eq:stability-1}).

These eigenvalues are interpreted as the double ionization and double
electron attachment energies in a molecular system, i.e.
\begin{equation}
\omega_{n}^{N+2}=E_{n}^{N+2}-E_{0}^{N}-2\nu,\label{eq:particleExcitationEigenvalue-1}
\end{equation}
for $N+2$ excitation energies, and 
\begin{equation}
\omega_{n}^{N-2}=E_{0}^{N}-E_{n}^{N-2}-2\nu,\label{eq:holeExcitationEigenvalue-1}
\end{equation}
or the $N-2$ excitation energies. With the eigenvalue interpretation
of Eqs. (\ref{eq:particleExcitationEigenvalue-1})-(\ref{eq:holeExcitationEigenvalue-1}),
an unstable pp-RPA equation violates the energetic convexity condition\cite{Parr89}.
It has not been proved that such stability is intrinsic for a self-consistent
solution of a Hartree-Fock or Kohn-Sham/generalized Kohn-Sham molecular
system, but in practice unstable solutions have never been encountered
for molecular systems so far in Ref.\cite{vanAggelen13} and in present
work.

The stability condition of the pp-RPA equation is equivalent to the
positive definiteness of the matrix $\mathbf{M}$. See Subsection
\ref{sec:The-relation-between} in Appendix for further details. The
positive definiteness as the stability criterion has been used in
Ref. \cite{Blaizot86}. 

With the whole spectrum of a stable pp-RPA equation, the pp-RPA correlation
energy can be expressed in several equivalent ways%
\cite{Note1}
\begin{align}
E_{\text{c}}^{\text{pp-RPA}} & =\sum_{m}\omega_{m}^{N+2}-\text{Tr}\mathbf{A}=-\sum_{n}\omega_{n}^{N-2}-\text{Tr}\mathbf{C}=\frac{1}{2}\sum_{n}|\omega_{n}|-\frac{1}{2}\text{Tr}\mathbf{M}.\label{eq:correlationEnergy}
\end{align}
 The precise value of $\nu$ is irrelevant for a stable pp-RPA equation
as long as
\[
\min_{m}(E_{m}^{N+2}-E_{0}^{N})>2\nu>\max_{n}(E_{0}^{N}-E_{n}^{N-2}),
\]
as they cancel out in the correlation energy expression. Yet a proper
chemical potential can categorize $\mathbf{M}$ to be positive definite,
an equivalent condition of the stability.

\section{Proof of the equivalence of pp-RPA and ladder-CCD\label{sec:Proof-of-Equivalence}}

The CCD ansatz, the simplest method in the coupled cluster family,
expresses the wavefunction as
\begin{equation}
|\text{CCD}\rangle=e^{\hat{T}_{2}}|\Phi_{0}\rangle,\label{eq:CCDAnsatz}
\end{equation}
where $|\Phi_{0}\rangle$ is a single Slater determinant, and $\hat{T}_{2}$
is the two-body cluster operator
\begin{equation}
\hat{T}_{2}=\frac{1}{2!}\sum_{ijab}t_{ij}^{ab}\hat{a}^{\dagger}\hat{i}\hat{b}^{\dagger}\hat{j}=\sum_{ijab}^{i>j,a>b}t_{ij}^{ab}\hat{a}^{\dagger}\hat{i}\hat{b}^{\dagger}\hat{j},\label{eq:clusterOperator}
\end{equation}
where $\hat{a}^{\dagger},\hat{i}$ are the creation and annihilation
operators for spin orbital $a$ and $i$, respectively and $t_{ij}^{ab}$
the double excitation amplitudes, having the the symmetry
\begin{equation}
t_{ij}^{ab}=-t_{ji}^{ab}=-t_{ij}^{ba}=t_{ji}^{ba}.\label{eq:amplitudeSymmetry}
\end{equation}

The correlation energy is expressed in terms of the amplitudes through
the energy equation
\begin{equation}
E_{\text{c}}^{\text{CCD}}=\sum_{ijab}^{i>j,a>b}\langle ij||ab\rangle t_{ij}^{ab},\label{eq:energyEquation}
\end{equation}
while the amplitudes $t_{ij}^{ab}$ are solved for by the CCD amplitude
equation,
\begin{align}
(\epsilon_{i}+\epsilon_{j}-\epsilon_{a}-\epsilon_{b})t_{ij}^{ab} & =\langle ab||ij\rangle+\frac{1}{2}\sum_{cd}\langle ab||cd\rangle t_{ij}^{cd}+\frac{1}{2}\sum_{kl}\langle ij||kl\rangle t_{kl}^{ab}\nonumber \\
 & \quad-\sum_{kc}(\langle bk||cj\rangle t_{ik}^{ac}-\langle bk||ci\rangle t_{jk}^{ac}-\langle ak||cj\rangle t_{ik}^{bc}+\langle ak||ci\rangle t_{jk}^{bc})\nonumber \\
 & \quad+\sum_{klcd}\langle kl||cd\rangle[\frac{1}{4}t_{ij}^{cd}t_{kl}^{ab}-\frac{1}{2}(t_{ij}^{ac}t_{kl}^{bd}+t_{ij}^{bd}t_{kl}^{ac})-\frac{1}{2}(t_{ik}^{ab}t_{jl}^{cd}+t_{ik}^{cd}t_{jl}^{ab})+(t_{ik}^{ac}t_{jl}^{bd}+t_{ik}^{bd}t_{jl}^{ac})].\label{eq:amplitudeEquation}
\end{align}
Refer to Ref. \cite{Shavitt09} for details of the CCD equations. 

By allowing only particle-hole summations in Eq. (\ref{eq:amplitudeEquation}),
Scuseria \emph{et al.\cite{Scuseria08231101}} have shown that the
amplitude equation reduces to the ph-RPA equation with exchange, i.e.,
the time-dependent Hartree-Fock (TDHF) equation. Further eliminating
the exchange term in the two-electron integral yields the conventional
direct ph-RPA. Similarly, if we allow only summations of particle
pairs and hole pairs, Eq. (\ref{eq:amplitudeEquation}) becomes
\begin{align}
 & \sum_{kl}(\epsilon_{k}+\epsilon_{l})t_{kl}^{ab}\delta_{ki}\delta_{jl}-\sum_{cd}(\epsilon_{c}+\epsilon_{d})t_{ij}^{cd}\delta_{ac}\delta_{bd}\nonumber \\
= & \langle ab||ij\rangle+\frac{1}{2}\sum_{cd}\langle ab||cd\rangle t_{ij}^{cd}+\frac{1}{2}\sum_{kl}\langle ij||kl\rangle t_{kl}^{ab}+\frac{1}{4}\sum_{kl,cd}t_{kl}^{ab}\langle kl||cd\rangle t_{ij}^{cd}.\label{eq:ladderCCD}
\end{align}
We refer to this restricted CCD as ladder-CCD, due to their inclusion
of only ladder diagrams in the correlation energy. By utilizing the
antisymmetry of the two-electron integrals $\langle uv||st\rangle=-\langle uv||ts\rangle$,
Eq. (\ref{eq:ladderCCD}) can be rearranged as
\begin{equation}
\sum_{cd}^{c>d}A_{ab,cd}t_{ij}^{cd}+\sum_{kl}^{k>l}C_{ij,kl}t_{kl}^{ab}+B_{ab,ij}+\sum_{kl,cd}^{k>l,c>d}t_{kl}^{ab}B_{cd,kl}^{*}t_{ij}^{cd}=0,\label{eq:ladderCCDRearrange}
\end{equation}
with $A$, $B$, and $C$ defined in Eqs. (\ref{eq:Adef})-(\ref{eq:Bdef}).
Denoting the amplitude as a matrix $T_{ab,ij}=t_{ij}^{ab},$ Eq. (\ref{eq:ladderCCDRearrange})
results in an algebraic matrix equation
\begin{equation}
\mathbf{AT}+\mathbf{TC}+\mathbf{B}+\mathbf{TB}^{\dagger}\mathbf{T}=0.\label{eq:ladderCCDMatrix}
\end{equation}
Now, we will show that the pp-RPA equation of Eq. (\ref{eq:ppRPA})
is equivalent to the ladder-CCD amplitude equation under the assumption
that the pp-RPA equation is stable. 

The pp-RPA equation for only the $N+2$ excitations reads,
\begin{equation}
\left[\begin{array}{cc}
\mathbf{A} & \mathbf{B}\\
\mathbf{B}^{\dagger} & \mathbf{C}
\end{array}\right]\left[\begin{array}{c}
\mathbf{X}\\
\mathbf{Y}
\end{array}\right]=\left[\begin{array}{cc}
\mathbf{I} & \mathbf{0}\\
\mathbf{0} & -\mathbf{I}
\end{array}\right]\left[\begin{array}{c}
\mathbf{X}\\
\mathbf{Y}
\end{array}\right]\bm{\omega}^{N+2},\label{eq:ppppRPA}
\end{equation}
where $\dim\mathbf{X}=N_{p}\times N_{p}$, $\dim\mathbf{Y}=N_{h}\times N_{p}$,
and $\text{dim}\bm{\omega}^{N+2}=N_{p}\times N_{p}$. Multiplying
$\mathbf{X}^{-1}$ from the right on Eq. (\ref{eq:ppppRPA}) gives
\begin{equation}
\left[\begin{array}{cc}
\mathbf{A} & \mathbf{B}\\
\mathbf{B}^{\dagger} & \mathbf{C}
\end{array}\right]\left[\begin{array}{c}
\mathbf{I}\\
\tilde{\mathbf{T}}^{\dagger}
\end{array}\right]=\left[\begin{array}{cc}
\mathbf{I} & \mathbf{0}\\
\mathbf{0} & -\mathbf{I}
\end{array}\right]\left[\begin{array}{c}
\mathbf{I}\\
\tilde{\mathbf{T}}^{\dagger}
\end{array}\right]\mathbf{R},\label{eq:transformedppRPA}
\end{equation}
where 
\begin{equation}
\tilde{\mathbf{T}}=(\mathbf{YX}^{-1})^{\dagger},\label{eq:2ndAmplitude}
\end{equation}
and
\begin{equation}
\mathbf{R}=\mathbf{X}\bm{\omega}^{N+2}\mathbf{X}^{-1}.\label{eq:R}
\end{equation}
The invertibility of $\mathbf{X}$ is guaranteed by a stable pp-RPA
equation. See Subsection \ref{sec:The-invertability-of} in Appendix
for the detailed proof. Multiplying $[\tilde{\mathbf{T}}^{\dagger}\ \mathbf{1}]$
from the left to Eq. (\ref{eq:transformedppRPA}) results in
\begin{equation}
\tilde{\mathbf{T}}^{\dagger}\mathbf{A}+\tilde{\mathbf{T}}^{\dagger}\mathbf{B}\tilde{\mathbf{T}}^{\dagger}+\mathbf{B}^{\dagger}+\mathbf{C}\tilde{\mathbf{T}}^{\dagger}=0.\label{eq:ccladderCCD}
\end{equation}
Comparing Eq. (\ref{eq:ladderCCDMatrix}) and Eq. (\ref{eq:ccladderCCD}),
we infer that $\mathbf{T}=\tilde{\mathbf{T}}$. 

The particle-particle block of Eq. (\ref{eq:transformedppRPA}) gives
\begin{equation}
\mathbf{A}+\mathbf{B}\mathbf{T}^{\dagger}=\mathbf{R}.\label{eq:particleBlockMatrix}
\end{equation}
Then, the ladder-CCD correlation energy of Eq. (\ref{eq:energyEquation})
can be expressed as
\begin{equation}
E_{\mathbf{c}}^{\mathbf{ladder-CCD}}=\text{Tr}(\mathbf{B}^{\dagger}\mathbf{T})=[\text{Tr}(\mathbf{R}-\mathbf{A})]^{*}=\sum_{m}\omega_{m}^{N+2}-\text{Tr}\mathbf{A},\label{eq:lCCD=00003DppRPA}
\end{equation}
which is identical to the pp-RPA correlation energy in Eq. (\ref{eq:correlationEnergy}).
From Eqs. (\ref{eq:ladderCCD})-(\ref{eq:ladderCCDMatrix}), it is
also clear that the chemical potential has no contribution because
they cancel each other in the CCD equations through $\mathbf{AT}+\mathbf{TC}$.

Alternatively, one can also derive the equivalence using the $N-2$
excitation eigenvectors with similar techniques. The resulting amplitude
will be the same, while the correlation energy expression will be
the second equation in Eq. (\ref{eq:correlationEnergy}).

In conclusion, the correlation energy from pp-RPA is equivalent to
that of ladder-CCD, assuming that the pp-RPA equation is stable. The
exponential wavefunction of Eq. (\ref{eq:CCDAnsatz}) with exponent
of Eq. (\ref{eq:2ndAmplitude}) has been proposed in Ref. \cite{Ring04},
together with a similar form for ph-RPA, however without exploring
their connection to the form of truncated CCD.

\section{Numerical demonstrations}

All coupled cluster and Møller\textendash{}Plesset perturbation theory
(MP2) computations reported herein are performed in a locally modified
version of CFOUR\cite{CFour}, while pp-RPA is performed with QM4D\cite{qm4d}.

Truncating the CCD equations to include only the ladder diagrams (Eq.
(\ref{eq:ladderCCD})) can be seen as a small modification of the
CCD equations or a small extension of the linearized CCD, also known
as CEPA(0) or D-MBPT($\infty$)\cite{Shavitt09}, amplitude equations.
Note that the computationally most expensive term of coupled-cluster
single and double (CCSD), scaling as $N_{\text{occ}}^{2}N_{\text{vir}}^{4}$,
is the major part of the term quadratic in the amplitudes of Eq. (\ref{eq:ladderCCD}).
In terms of efficiency, the matrix multiplications necessary for solving
the non-linear system of equations in standard coupled cluster algorithms
are traded against the diagonalization in the pp-RPA algorithm, which,
at the non-optimized stage of the code,\cite{qm4d} is significantly
slower than solving the non-linear equations. However, the diagonalization
has the indisputable advantage that the solution is unique, whereas
the non-linear coupled cluster equations have multiple minima (most
of them lacking any physical meaning), without an \textit{a priori}
guarantee or check that the ``correct'' solution is found.\cite{Shavitt09}

All computations are carried out in the unrestricted Hartree-Fock
(UHF) framework, but without breaking space symmetry. The correlation
consistent basis sets of Dunning and coworkers\cite{Dunning891007,Woon931358}
have been applied with cartesian d- and f- atomic-orbitals. The ladder-CCD
amplitudes are found to converge essentially as fast (or with a couple
of iterations less) than the corresponding CCSD equations.

All total energies of ladder-CCD and pp-RPA (see Table \ref{tab:Total-energies-1})
agree exceedingly well, the largest difference being $10^{-5}$ Hartree,
which is on the same order of magnitude as the difference in nuclear
repulsion energy between the two programs and can have its origin
in, e.g., integral screening (SCF and CC iteration convergence has
been checked carefully). In terms of correlation energy, ladder-CCD
captures between 43\% (Be) to 80\% (Ne) of CCSD, while the full CCD
energy recovers about 99\%. Note that MP2 has min and max values of
70\% and 99\% for the same systems. Furthermore, changing to a DFT
reference%
\cite{Note2}
 leads to an increased (in absolute terms) correlation energy, with
min/max values reaching 51(54)\% and 92 (95)\% for B3LYP\cite{Becke935648,Lee88785}
(PBE\cite{Perdew963865}) orbitals.

\begin{center}
\begin{table}
\caption{Total energies of various methods. Geometries are taken from the G3
set\cite{Curtiss007374,Curtiss05124107}. The basis set is cc-pVTZ,
except for benzene where cc-pVDZ is applied. All energies are in Hartree\label{tab:Total-energies-1}}

\centering{}%
\begin{tabular}{|c||r@{\extracolsep{0pt}.}l|r@{\extracolsep{0pt}.}l|r@{\extracolsep{0pt}.}l|r@{\extracolsep{0pt}.}l|r@{\extracolsep{0pt}.}l|r@{\extracolsep{0pt}.}l|r@{\extracolsep{0pt}.}l|r@{\extracolsep{0pt}.}l|}
\hline 
 & \multicolumn{2}{c|}{{\footnotesize{HF}}} & \multicolumn{2}{c|}{{\footnotesize{pp-RPA@HF}}} & \multicolumn{2}{c|}{{\footnotesize{ladder-CCD}}} & \multicolumn{2}{c|}{{\footnotesize{pp-RPA@PBE}}} & \multicolumn{2}{c|}{{\footnotesize{pp-RPA@B3LYP}}} & \multicolumn{2}{c|}{{\footnotesize{MP2}}} & \multicolumn{2}{c|}{{\footnotesize{CCD}}} & \multicolumn{2}{c|}{{\footnotesize{CCSD}}}\tabularnewline
\hline 
\hline 
{\footnotesize{He}} & {\footnotesize{-2}}&{\footnotesize{861154}} & {\footnotesize{-2}}&{\footnotesize{885608}} & {\footnotesize{-2}}&{\footnotesize{885608}} & {\footnotesize{-2}}&{\footnotesize{889343}} & {\footnotesize{-2}}&{\footnotesize{888504}} & {\footnotesize{-2}}&{\footnotesize{894441}} & {\footnotesize{-2}}&{\footnotesize{900328}} & {\footnotesize{-2}}&{\footnotesize{900351}}\tabularnewline
\hline 
{\footnotesize{Li}} & {\footnotesize{-7}}&{\footnotesize{432706}} & {\footnotesize{-7}}&{\footnotesize{443903}} & {\footnotesize{-7}}&{\footnotesize{443903}} & {\footnotesize{-7}}&{\footnotesize{444664}} & {\footnotesize{-7}}&{\footnotesize{444450}} & {\footnotesize{-7}}&{\footnotesize{446781}} & {\footnotesize{-7}}&{\footnotesize{449184}} & {\footnotesize{-7}}&{\footnotesize{449243}}\tabularnewline
\hline 
{\footnotesize{Be}} & {\footnotesize{-14}}&{\footnotesize{572875}} & {\footnotesize{-14}}&{\footnotesize{598923}} & {\footnotesize{-14}}&{\footnotesize{598923}} & {\footnotesize{-14}}&{\footnotesize{605231}} & {\footnotesize{-14}}&{\footnotesize{603533}} & {\footnotesize{-14}}&{\footnotesize{614751}} & {\footnotesize{-14}}&{\footnotesize{632242}} & {\footnotesize{-14}}&{\footnotesize{632817}}\tabularnewline
\hline 
{\footnotesize{B}} & {\footnotesize{-24}}&{\footnotesize{532104}} & {\footnotesize{-24}}&{\footnotesize{566435}} & {\footnotesize{-24}}&{\footnotesize{566436}} & {\footnotesize{-24}}&{\footnotesize{575674}} & {\footnotesize{-24}}&{\footnotesize{573063}} & {\footnotesize{-24}}&{\footnotesize{584950}} & {\footnotesize{-24}}&{\footnotesize{604746}} & {\footnotesize{-24}}&{\footnotesize{605490}}\tabularnewline
\hline 
{\footnotesize{C}} & {\footnotesize{-37}}&{\footnotesize{691663}} & {\footnotesize{-37}}&{\footnotesize{746778}} & {\footnotesize{-37}}&{\footnotesize{746778}} & {\footnotesize{-37}}&{\footnotesize{760145}} & {\footnotesize{-37}}&{\footnotesize{756583}} & {\footnotesize{-37}}&{\footnotesize{769564}} & {\footnotesize{-37}}&{\footnotesize{789208}} & {\footnotesize{-37}}&{\footnotesize{789809}}\tabularnewline
\hline 
{\footnotesize{N}} & {\footnotesize{-54}}&{\footnotesize{400883}} & {\footnotesize{-54}}&{\footnotesize{482916}} & {\footnotesize{-54}}&{\footnotesize{482916}} & {\footnotesize{-54}}&{\footnotesize{500883}} & {\footnotesize{-54}}&{\footnotesize{496235}} & {\footnotesize{-54}}&{\footnotesize{509992}} & {\footnotesize{-54}}&{\footnotesize{525553}} & {\footnotesize{-54}}&{\footnotesize{525893}}\tabularnewline
\hline 
{\footnotesize{O}} & {\footnotesize{-74}}&{\footnotesize{811910}} & {\footnotesize{-74}}&{\footnotesize{933839}} & {\footnotesize{-74}}&{\footnotesize{933839}} & {\footnotesize{-74}}&{\footnotesize{959853}} & {\footnotesize{-74}}&{\footnotesize{953384}} & {\footnotesize{-74}}&{\footnotesize{969918}} & {\footnotesize{-74}}&{\footnotesize{985506}} & {\footnotesize{-74}}&{\footnotesize{986128}}\tabularnewline
\hline 
{\footnotesize{F}} & {\footnotesize{-99}}&{\footnotesize{405657}} & {\footnotesize{-99}}&{\footnotesize{576884}} & {\footnotesize{-99}}&{\footnotesize{576884}} & {\footnotesize{-99}}&{\footnotesize{611587}} & {\footnotesize{-99}}&{\footnotesize{603292}} & {\footnotesize{-99}}&{\footnotesize{622736}} & {\footnotesize{-99}}&{\footnotesize{633484}} & {\footnotesize{-99}}&{\footnotesize{634177}}\tabularnewline
\hline 
{\footnotesize{Ne}} & {\footnotesize{-128}}&{\footnotesize{532010}} & {\footnotesize{-128}}&{\footnotesize{760771}} & {\footnotesize{-128}}&{\footnotesize{760771}} & {\footnotesize{-128}}&{\footnotesize{804849}} & {\footnotesize{-128}}&{\footnotesize{794546}} & {\footnotesize{-128}}&{\footnotesize{816523}} & {\footnotesize{-128}}&{\footnotesize{817814}} & {\footnotesize{-128}}&{\footnotesize{818536}}\tabularnewline
\hline 
{\footnotesize{CH$_{4}$}} & {\footnotesize{-40}}&{\footnotesize{213408}} & {\footnotesize{-40}}&{\footnotesize{372051}} & {\footnotesize{-40}}&{\footnotesize{372054}} & {\footnotesize{-40}}&{\footnotesize{411910}} & {\footnotesize{-40}}&{\footnotesize{402169}} & {\footnotesize{-40}}&{\footnotesize{432266}} & {\footnotesize{-40}}&{\footnotesize{452031}} & {\footnotesize{-40}}&{\footnotesize{452991}}\tabularnewline
\hline 
{\footnotesize{H$_{2}$O}} & {\footnotesize{-76}}&{\footnotesize{056687}} & {\footnotesize{-76}}&{\footnotesize{266046}} & {\footnotesize{-76}}&{\footnotesize{266049}} & {\footnotesize{-76}}&{\footnotesize{318304}} & {\footnotesize{-76}}&{\footnotesize{305731}} & {\footnotesize{-76}}&{\footnotesize{336459}} & {\footnotesize{-76}}&{\footnotesize{340863}} & {\footnotesize{-76}}&{\footnotesize{342084}}\tabularnewline
\hline 
{\footnotesize{NH$_{3}$}} & {\footnotesize{-56}}&{\footnotesize{217964}} & {\footnotesize{-56}}&{\footnotesize{404439}} & {\footnotesize{-56}}&{\footnotesize{404440}} & {\footnotesize{-56}}&{\footnotesize{452289}} & {\footnotesize{-56}}&{\footnotesize{440556}} & {\footnotesize{-56}}&{\footnotesize{471921}} & {\footnotesize{-56}}&{\footnotesize{483441}} & {\footnotesize{-56}}&{\footnotesize{484474}}\tabularnewline
\hline 
{\footnotesize{CH$_{2}$O}} & {\footnotesize{-113}}&{\footnotesize{910280}} & {\footnotesize{-114}}&{\footnotesize{227562}} & {\footnotesize{-114}}&{\footnotesize{227552}} & {\footnotesize{-114}}&{\footnotesize{313824}} & {\footnotesize{-114}}&{\footnotesize{293495}} & {\footnotesize{-114}}&{\footnotesize{341669}} & {\footnotesize{-114}}&{\footnotesize{347547}} & {\footnotesize{-114}}&{\footnotesize{351726}}\tabularnewline
\hline 
{\footnotesize{C$_{6}$H$_{6}$}} & {\footnotesize{-230}}&{\footnotesize{722701}} & {\footnotesize{-231}}&{\footnotesize{315273}} & {\footnotesize{-231}}&{\footnotesize{315273}} & {\footnotesize{-231}}&{\footnotesize{508132}} & {\footnotesize{-231}}&{\footnotesize{460711}} & {\footnotesize{-231}}&{\footnotesize{540504}} & {\footnotesize{-231}}&{\footnotesize{571751}} & {\footnotesize{-231}}&{\footnotesize{577366}}\tabularnewline
\hline 
\end{tabular}
\end{table}

\par\end{center}

As a graphical illustration, Figure \ref{fig:The-potential-energy}
shows the case of a dissociating cationic dimer (Ne$_{2}^{+}$), a
typical probe for (de)localization error. Again, the total energies
of ladder-CCD and pp-RPA are identical to numerical precision (considering
the two very different algorithms and programs), but not in very good
agreement with CCSD. To further investigate the (de)localization error\cite{Mori-Sanchez0814},
Figure \ref{fig:The-binding-curve} shows the binding energy with
respect to the separated fragments. The binding energy of ladder-CCD
is in fairly good agreement with CCSD and only a small ``bump''
is observed somewhere between 3 and 4 Å, revealing that the missing
absolute correlation energies in ladder-CCD compared to CCSD are almost
irrelevant for the binding energy. The localization error of HF is
over-corrected by MP2, but increasing the correlation treatment to
the coupled cluster level improves the dissociation limit further,
leading to the previously reported\cite{vanAggelen13} negligible
fractional charge error.

\begin{center}
\begin{figure}
\begin{centering}
\subfloat[The potential energy surface of Ne$_{2}^{+}$\label{fig:The-potential-energy}]{\begin{centering}
\includegraphics[width=0.7\textwidth, natwidth=160.165mm, natheight=88.1789mm]{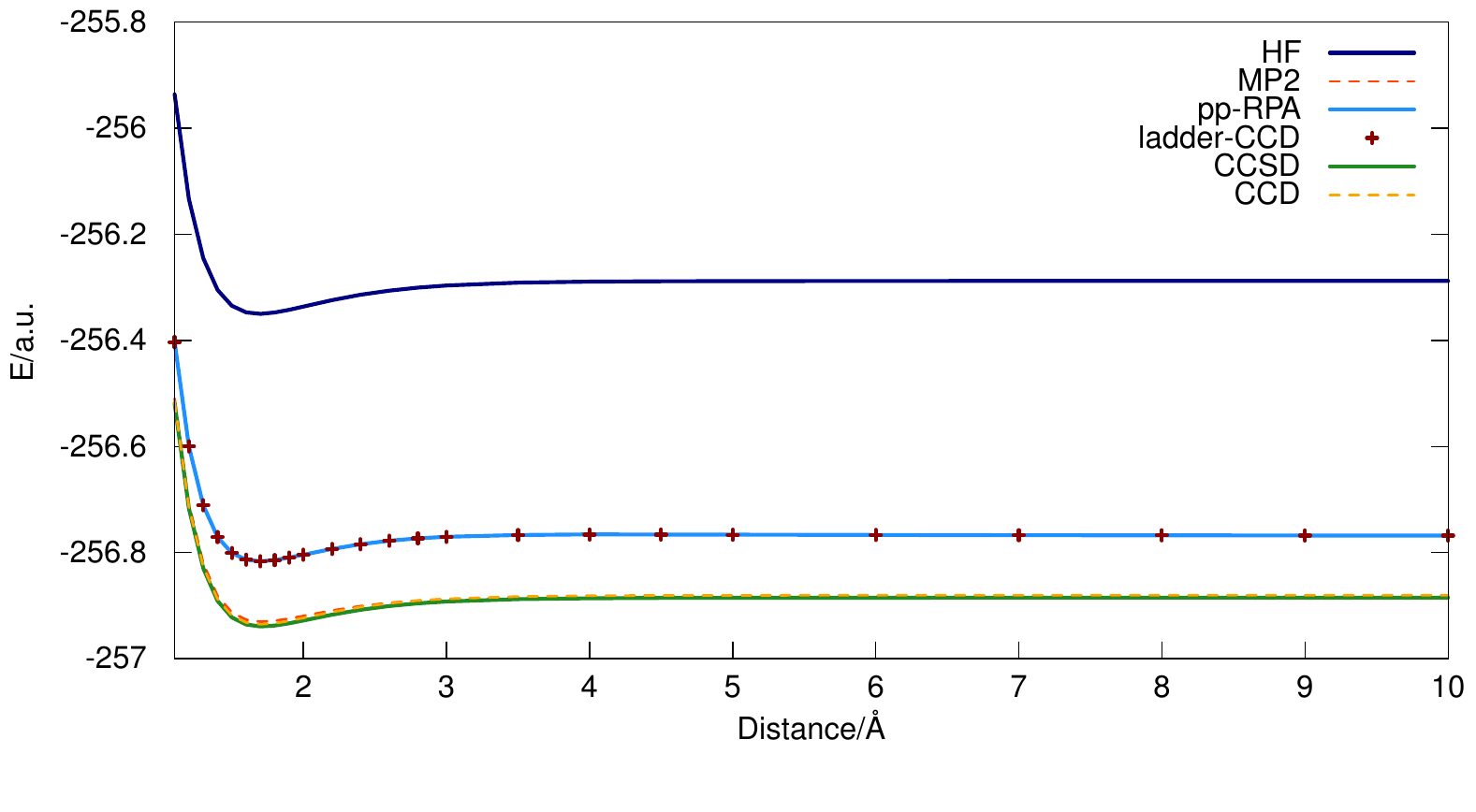}
\par\end{centering}

}
\par\end{centering}

\centering{}\subfloat[The binding curve of Ne$_{2}^{+}$ (with respect to Ne and Ne$^{+}$)\label{fig:The-binding-curve}]{\centering{}\includegraphics[width=0.7\textwidth,  natwidth=160.165mm, natheight=88.1789mm]{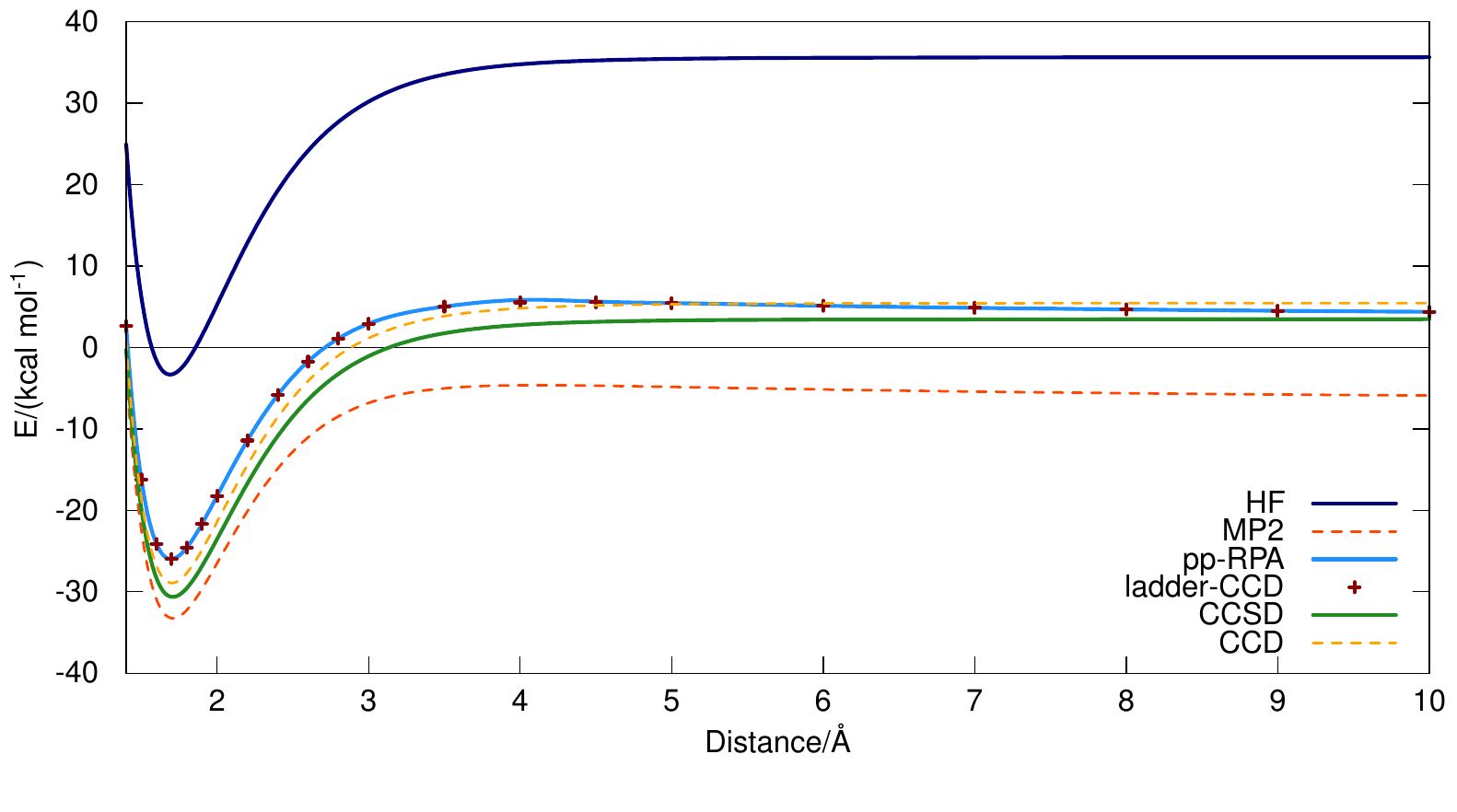}}\caption{The potential energy surface (a) and the binding curve (b) of Ne$_{2}^{+}$
of various methods with basis set aug-cc-pVTZ.\label{fig:ENe2+-1}
The total energies of pp-RPA are substantially overestimated (a),
since the correlation energy of the ladder diagrams is not very well
balanced (MP2 total energies are, on the scale of the figure, indistinguishable
from CCD, and pp-RPA is correct through second order\cite{vanAggelen13}).
However, the binding energy (b) reveals that the missing correlation
energy cancels almost perfectly out, yielding a pp-RPA binding energy
curve very close to CCD, while MP2 deviates from CCSD in the other
direction (overbinding).}
\end{figure}

\par\end{center}

Similarly to the binding energy of Ne$_{2}^{+}$, the atomization
energies (Table \ref{tab:EAt}) illustrate that the correlation energy
missing in ladder-CCD largely cancels out when computing reaction
energies. For the five molecules considered, ladder-CCD provides 73\%
(CH\textsubscript{4}) to 91\% (C\textsubscript{6}H\textsubscript{6})
of the correction between the HF and CCSD atomization energy. This
is to be compared with MP2 which recovers between 100\% and 122\%.
The range for pp-RPA@B3LYP (pp-RPA@PBE) is, with 88 (93)\% for methane
to 115 (122)\% for benzene somewhat larger. In summary, the numerical
analysis shows that ladder-CCD and pp-RPA are equivalent and that
the chemically relevant correlation contributions missing in ladder-CCD
compared to CCSD are relatively small. An efficient pp-RPA implementation
has, therefore, the potential to become a valuable electronic structure
theory.

\begin{center}
\begin{table}
\caption{Atomization energies (in kcal mol\textsuperscript{-1}) of various
methods. Geometries are taken from the G3 set\cite{Curtiss007374,Curtiss05124107}.
The basis set is cc-pVTZ, except for benzene where cc-pVDZ is applied.
The mean absolute deviation (MAD) is with respect to CCSD.\label{tab:EAt}}

\centering{}%
\begin{tabular}{|c||r@{\extracolsep{0pt}.}l|r@{\extracolsep{0pt}.}l|r@{\extracolsep{0pt}.}l|r@{\extracolsep{0pt}.}l|r@{\extracolsep{0pt}.}l|r@{\extracolsep{0pt}.}l|r@{\extracolsep{0pt}.}l|r@{\extracolsep{0pt}.}l|}
\hline 
 & \multicolumn{2}{c|}{HF} & \multicolumn{2}{c|}{pp-RPA@HF} & \multicolumn{2}{c|}{ladder-CCD} & \multicolumn{2}{c|}{pp-RPA@PBE} & \multicolumn{2}{c|}{pp-RPA@B3LYP} & \multicolumn{2}{c|}{MP2} & \multicolumn{2}{c|}{CCD} & \multicolumn{2}{c|}{CCSD}\tabularnewline
\hline 
\hline 
CH$_{4}$ & 327&88 & 392&84 & 392&84 & 410&65 & 406&40 & 416&33 & 416&40 & 416&63\tabularnewline
\hline 
H$_{2}$O & 153&84 & 208&70 & 208&70 & 225&76 & 221&74 & 230&25 & 223&23 & 223&60\tabularnewline
\hline 
NH$_{3}$ & 199&33 & 264&87 & 264&87 & 284&51 & 279&78 & 290&22 & 287&69 & 288&12\tabularnewline
\hline 
CH$_{2}$O & 255&45 & 343&45 & 343&45 & 373&46 & 366&81 & 378&12 & 359&70 & 361&55\tabularnewline
\hline 
C$_{6}$H$_{6}$ & 1008&42 & 1237&59 & 1237&59 & 1315&30 & 1296&70 & 1315&24 & 1258&08 & 1259&81\tabularnewline
\hline 
\hline 
MAD & 120&96  & 20&45  & 20&45  & 15&83  &  12&52 & 16&21  & 0&92  & \multicolumn{2}{c|}{--}\tabularnewline
\hline 
\end{tabular}
\end{table}

\par\end{center}

\section{Conclusions}

The equivalence of the pp-RPA correlation energy and the ladder-CCD
approach has been analytically proved, with the assumption that the
pp-RPA equation is stable, and numerically demonstrated. The numerical
assessment suggests that the missing correlation in pp-RPA is favorably
canceled out in reaction energies. The ladder-CCD perspective of the
pp-RPA correlation energy purveys a concrete wavefunction of the ground
state, which makes the study of its ground and excited state properties
straight forward.

\section*{Acknowledgment}

Support from the Office of Naval Research (ONR) (N00014-09-1-0576),
and the National Science Foundation (NSF) (CHE-09-11119) is gratefully
appreciated. D.P. has also been supported by the William Krigbaum
and Marcus Hobbs Fellowship from Duke University. S.N.S. acknowledges
the Swiss NSF fellowship PBELP2\_143559. H.v.A. appreciates the support
form the FWO-Flanders (Scientific Research Fund Flanders).

%

\appendix

\section{Mathematical analysis of the pp-RPA equation}

\subsection{The zero signature of an eigenvector with an imaginary eigenvalue\label{sec:The-zero-signature}}

For an eigenvalue $\omega_{n}$ and eigenvector $\mathbf{z}_{n}$,
we have
\begin{equation}
\mathbf{Mz}_{n}=\omega_{n}\mathbf{Wz}_{n}.\label{eq:compactEigen}
\end{equation}
The Hermitian conjugate of Eq. (\ref{eq:compactEigen}) becomes
\begin{equation}
\mathbf{z}_{n}^{\dagger}\mathbf{M}=\omega_{n}^{*}\mathbf{z}_{n}^{\dagger}\mathbf{W}.\label{eq:HermitianConjugate}
\end{equation}
Multiplying $\mathbf{z}_{n}^{\dagger}$ to the left of Eq. (\ref{eq:compactEigen})
and $\mathbf{z}_{n}$ to the right of Eq. (\ref{eq:HermitianConjugate}),
we have 
\[
\mathbf{z}_{n}^{\dagger}\mathbf{Mz}_{n}=\omega_{n}\mathbf{z}_{n}^{\dagger}\mathbf{Wz}_{n}=\omega_{n}^{*}\mathbf{z}_{n}^{\dagger}\mathbf{Wz}_{n}.
\]
Therefore
\begin{equation}
(\omega_{n}-\omega_{n}^{*})(\mathbf{z}_{n}^{\dagger}\mathbf{Wz}_{n})=0.\label{eq:imaginaryEigenvalue}
\end{equation}

For an imaginary eigenvalue $\omega_{n}\ne\omega_{n}^{*}$, the signature
$\mathbf{z}_{n}^{\dagger}\mathbf{Wz}_{n}=0$.

\subsection{The orthonormalization of eigenvectors with all real eigenvalues\label{sec:orthonormalization}}

Using the same approach in Subsection \ref{sec:The-zero-signature}
in Appendix but with two different eigenvalues and eigenvectors, we
have
\[
\mathbf{z}_{n}^{\dagger}\mathbf{Mz}_{m}=\omega_{m}\mathbf{z}_{n}^{\dagger}\mathbf{W}\mathbf{z}_{m}=\omega_{n}^{*}\mathbf{z}_{n}^{\dagger}\mathbf{Wz}_{m},
\]
and
\begin{equation}
(\omega_{m}-\omega_{n}^{*})(\mathbf{z}_{n}^{\dagger}\mathbf{Wz}_{m})=0.\label{eq:orthogonal}
\end{equation}
Therefore, when two real eigenvalues are different ($\omega_{m}\ne\omega_{n}^{*}$),
the two eigenvectors are orthogonal under the metric $\mathbf{W}$
($\mathbf{z}_{n}^{\dagger}\mathbf{Wz}_{m}=0$). Since linear combination
of eigenvectors of a degenerate eigenvalue stays in the same eigenspace,
we can choose the eigenvectors of a degenerate eigenvalue to orthogonal
to each other within the eigenspace. When all eigenvalues are real,
eigenvectors can, therefore, be chosen to be orthogonalized under
the metric $\mathbf{W}$. For a diagonalizable pp-RPA equation with
all real eigenvalues, $\mathbf{z}_{n}^{\dagger}\mathbf{Wz}_{n}$ should
not be zero, otherwise we have $\mathbf{z}_{n}^{\dagger}\mathbf{WZ}=0$,
which indicates the eigenvector matrix is rank-deficit, which contradicts
with the diagonalizability assumption. Therefore, the signatures of
eigenvectors are all nonzero for a diagonalizable pp-RPA equation
with all real eigenvalues. The resulting orthonormalization can be
written as
\begin{equation}
\mathbf{Z}^{\dagger}\mathbf{WZ}=\Lambda,\label{eq:Lambda}
\end{equation}
where $\Lambda$ is a diagonal matrix with only $\pm1$ diagonal elements.
According to Sylvester's law of inertia\cite{Horn90}, $\mathbf{W}$
and $\Lambda$ share the same number of $+1$'s and $-1$'s. In another
word, there are $N_{pp}$ $N+2$ excitations and $N_{hh}$ $N-2$
excitations, according to the definition of $N\pm2$ excitations in
Sec. \ref{sec:The-ppRPA-equation}. We can further arrange the eigenvectors
such that eigenvectors with positive signatures stay in the left of
$\mathbf{Z}$, then finally we reach the normalization condition
\begin{equation}
\mathbf{Z}^{\dagger}\mathbf{WZ}=\mathbf{W}.\label{eq:FinalOrthonormalization}
\end{equation}

\subsection{The equivalence between stability and positive definiteness of $\mathbf{M}$\label{sec:The-relation-between}}

First we show that the stability condition of Eq. (\ref{eq:stability-1})
leads to the positive definiteness of $\mathbf{M}$.

From the stability of the pp-RPA equation (Eq. (\ref{eq:stability-1}))
and the normalization (Eq. (\ref{eq:eigenvectorNormalization})),
we have
\begin{align*}
\mathbf{c}^{\dagger}\mathbf{M}\mathbf{c} & =\sum_{mn}(\mathbf{z}_{m}c_{m})^{\dagger}\mathbf{M}(\mathbf{z}_{n}c_{n})\\
 & =\sum_{mn}c_{m}^{*}\mathbf{z}_{m}^{\dagger}\omega_{n}\mathbf{W}\mathbf{z}_{n}c_{n}\\
 & =\sum_{n}c_{m}^{*}\delta_{mn}W_{mn}\omega_{n}c_{n}\\
 & =\sum_{mn}c_{m}^{*}|\omega_{m}|\delta_{mn}c_{n}\\
 & =\sum_{m}|c_{m}|^{2}|\omega_{m}|>0,
\end{align*}
with an arbitrary nonzero column vector $\mathbf{c}$. Thus, $\mathbf{M}$
is positive definite for a pp-RPA equation.

Next, we show that the reverse is also true.

Given that $\mathbf{M}$ is positive definite, the pp-RPA equation
in the compact form reads 

\begin{equation}
\mathbf{M}\mathbf{z}_{n}=\omega_{n}\mathbf{Wz}_{n}.\label{eq:eigenvalues}
\end{equation}
Since $\mathbf{M}$ is positive definite, Eq. (\ref{eq:compact})
could be rewritten as
\[
\mathbf{L}^{\dagger}\mathbf{z}_{n}=\omega_{n}\mathbf{L}^{-1}\mathbf{W}\left(\mathbf{L}^{-1}\right)^{\dagger}\mathbf{L}^{\dagger}\mathbf{z}_{n},
\]
where $\mathbf{M}=\mathbf{LL}^{\dagger}$ is the Cholesky decomposition.
With $\tilde{\mathbf{z}}_{n}=\mathbf{L}^{\dagger}\mathbf{z}_{n}$
and $\tilde{\mathbf{W}}=\mathbf{L}^{-1}\mathbf{W}\left(\mathbf{L}^{-1}\right)^{\dagger},$
then the eigenvalue problem
\begin{equation}
\tilde{\mathbf{W}}\tilde{\mathbf{z}}_{n}=\tilde{\omega}_{n}\tilde{\mathbf{z}}_{n}\label{eq:diagonalizable}
\end{equation}
is diagonalizable with all real eigenvalues, since $\tilde{\mathbf{W}}^{\dagger}=\tilde{\mathbf{W}}$
by definition. Additionally, all eigenvalues of $\tilde{\mathbf{W}}$,
$\tilde{\omega}_{n}$'s, will be nonzero, since zero eigenvalue indicates
$\text{det}(\tilde{\mathbf{W}})=0$ which contradicts the definition
of $\tilde{\mathbf{W}}$. With orthonormalization of the eigenvectors
$\tilde{\mathbf{z}}_{n}^{\dagger}\tilde{\mathbf{z}}_{m}=\delta_{nm}|\tilde{\omega}_{n}|^{-1}$,
Eq. (\ref{eq:compact}) can be diagonalized with real eigenvalues
\begin{equation}
\omega_{n}=\tilde{\omega}_{n}^{-1},\label{eq:eigenvalueRelation}
\end{equation}
and eigenvector orthonormalization with the eigenvalue sign constraints
(the eigenvectors are arranged in the same way as in Subsection \ref{sec:orthonormalization}
in Appendix), 
\begin{equation}
\mathbf{z}_{n}^{\dagger}\mathbf{W}\mathbf{z}_{m}=\delta_{mn}\text{sign}(\omega_{m})=W_{nm}.\label{eq:subStability}
\end{equation}
Eq. (\ref{eq:subStability}) guarantees that the $\min_{n}\omega_{n}^{N+2}>0>\max_{m}\omega_{m}^{N-2}$.
Therefore, by definition, this pp-RPA equation is stable since all
the eigenvalues are real and the $N+2$ and $N-2$ excitation spectra
are nicely separated.

In summary, the stability condition of an pp-RPA equation is equivalent
to the positive definiteness of $\mathbf{M}$.

\subsection{The invertibility of $\mathbf{X}$ for a stable pp-RPA equation\label{sec:The-invertability-of}}

We now prove the invertibility of $\mathbf{X}$ in Sec. \ref{sec:Proof-of-Equivalence}.
According to Subsection \ref{sec:orthonormalization} in Appendix,
the eigenvalues of a stable pp-RPA equation are orthonormalized according
to 
\begin{equation}
\mathbf{Z}^{\dagger}\mathbf{WZ}=\mathbf{W}.\label{eq:a4normalization}
\end{equation}
For only $N+2$ excitation vectors,
\begin{equation}
\mathbf{Z}_{N+2}^{\dagger}\mathbf{WZ}_{N+2}=\mbox{\ensuremath{\mathbf{I}}},\label{eq:MatrixOrthonormalization}
\end{equation}
where 
\[
\mathbf{Z}_{N+2}=\left[\begin{array}{c}
\mathbf{X}\\
\mathbf{Y}
\end{array}\right],
\]
with $\mathbf{X}$ and $\mathbf{Y}$ the particle-particle and hole-hole
block of the $N+2$ excitation eigenvector matrices. Expanding Eq.
(\ref{eq:MatrixOrthonormalization}), we have
\begin{equation}
\mathbf{X}^{\dagger}\mathbf{X}-\mathbf{Y}^{\dagger}\mathbf{Y}=\mathbf{I}.\label{eq:ppBlock}
\end{equation}
Therefore, $\mathbf{X}^{\dagger}\mathbf{X}=\mathbf{I}+\mathbf{Y}^{\dagger}\mathbf{Y}$
is positive definite, and $\mathbf{X}$ is invertible, otherwise $\mathbf{X}^{\dagger}\mathbf{X}$
will not be positive definite.
\end{document}